\documentclass[aps,prl,superscriptaddress,showpacs,floatfix,twocolumn]{revtex4}

\usepackage{graphicx}   

\begin{document}

\title{Measurement of high-$p_{\rm T}$ Single Electrons from Heavy-Flavor Decays in $p+p$ Collisions at $\sqrt{s}$~=~200~GeV}

\newcommand{\abilene}{Abilene Christian University, Abilene, TX 79699, U.S.}
\newcommand{\banaras}{Department of Physics, Banaras Hindu University, Varanasi 221005, India}
\newcommand{\bnl}{Brookhaven National Laboratory, Upton, NY 11973-5000, U.S.}
\newcommand{\caucr}{University of California - Riverside, Riverside, CA 92521, U.S.}
\newcommand{\charlesczech}{Charles University, Ovocn\'{y} trh 5, Praha 1, 116 36, Prague, Czech Republic}
\newcommand{\ciae}{China Institute of Atomic Energy (CIAE), Beijing, People's Republic of China}
\newcommand{\cns}{Center for Nuclear Study, Graduate School of Science, University of Tokyo, 7-3-1 Hongo, Bunkyo, Tokyo 113-0033, Japan}
\newcommand{\colorado}{University of Colorado, Boulder, CO 80309, U.S.}
\newcommand{\columbia}{Columbia University, New York, NY 10027 and Nevis Laboratories, Irvington, NY 10533, U.S.}
\newcommand{\czechtech}{Czech Technical University, Zikova 4, 166 36 Prague 6, Czech Republic}
\newcommand{\dapnia}{Dapnia, CEA Saclay, F-91191, Gif-sur-Yvette, France}
\newcommand{\debrecen}{Debrecen University, H-4010 Debrecen, Egyetem t{\'e}r 1, Hungary}
\newcommand{\elte}{ELTE, E{\"o}tv{\"o}s Lor{\'a}nd University, H - 1117 Budapest, P{\'a}zm{\'a}ny P. s. 1/A, Hungary}
\newcommand{\fit}{Florida Institute of Technology, Melbourne, FL 32901, U.S.}
\newcommand{\fsu}{Florida State University, Tallahassee, FL 32306, U.S.}
\newcommand{\gsu}{Georgia State University, Atlanta, GA 30303, U.S.}
\newcommand{\hiroshima}{Hiroshima University, Kagamiyama, Higashi-Hiroshima 739-8526, Japan}
\newcommand{\ihepprot}{IHEP Protvino, State Research Center of Russian Federation, Institute for High Energy Physics, Protvino, 142281, Russia}
\newcommand{\illuiuc}{University of Illinois at Urbana-Champaign, Urbana, IL 61801, U.S.}
\newcommand{\instpasczech}{Institute of Physics, Academy of Sciences of the Czech Republic, Na Slovance 2, 182 21 Prague 8, Czech Republic}
\newcommand{\isu}{Iowa State University, Ames, IA 50011, U.S.}
\newcommand{\jinrdubna}{Joint Institute for Nuclear Research, 141980 Dubna, Moscow Region, Russia}
\newcommand{\kek}{KEK, High Energy Accelerator Research Organization, Tsukuba, Ibaraki 305-0801, Japan}
\newcommand{\kfki}{KFKI Research Institute for Particle and Nuclear Physics of the Hungarian Academy of Sciences (MTA KFKI RMKI), H-1525 Budapest 114, POBox 49, Budapest, Hungary}
\newcommand{\korea}{Korea University, Seoul, 136-701, Korea}
\newcommand{\kurchatov}{Russian Research Center ``Kurchatov Institute", Moscow, Russia}
\newcommand{\kyoto}{Kyoto University, Kyoto 606-8502, Japan}
\newcommand{\labllr}{Laboratoire Leprince-Ringuet, Ecole Polytechnique, CNRS-IN2P3, Route de Saclay, F-91128, Palaiseau, France}
\newcommand{\lawllnl}{Lawrence Livermore National Laboratory, Livermore, CA 94550, U.S.}
\newcommand{\losalamos}{Los Alamos National Laboratory, Los Alamos, NM 87545, U.S.}
\newcommand{\lpc}{LPC, Universit{\'e} Blaise Pascal, CNRS-IN2P3, Clermont-Fd, 63177 Aubiere Cedex, France}
\newcommand{\lund}{Department of Physics, Lund University, Box 118, SE-221 00 Lund, Sweden}
\newcommand{\muenster}{Institut f\"ur Kernphysik, University of Muenster, D-48149 Muenster, Germany}
\newcommand{\myongji}{Myongji University, Yongin, Kyonggido 449-728, Korea}
\newcommand{\nagasaki}{Nagasaki Institute of Applied Science, Nagasaki-shi, Nagasaki 851-0193, Japan}
\newcommand{\newmex}{University of New Mexico, Albuquerque, NM 87131, U.S. }
\newcommand{\nmsu}{New Mexico State University, Las Cruces, NM 88003, U.S.}
\newcommand{\ornl}{Oak Ridge National Laboratory, Oak Ridge, TN 37831, U.S.}
\newcommand{\orsay}{IPN-Orsay, Universite Paris Sud, CNRS-IN2P3, BP1, F-91406, Orsay, France}
\newcommand{\peking}{Peking University, Beijing, People's Republic of China}
\newcommand{\pnpi}{PNPI, Petersburg Nuclear Physics Institute, Gatchina, Leningrad region, 188300, Russia}
\newcommand{\riken}{RIKEN, The Institute of Physical and Chemical Research, Wako, Saitama 351-0198, Japan}
\newcommand{\rikjrbrc}{RIKEN BNL Research Center, Brookhaven National Laboratory, Upton, NY 11973-5000, U.S.}
\newcommand{\rikkyo}{Physics Department, Rikkyo University, 3-34-1 Nishi-Ikebukuro, Toshima, Tokyo 171-8501, Japan}
\newcommand{\saispbstu}{Saint Petersburg State Polytechnic University, St. Petersburg, Russia}
\newcommand{\saopaulo}{Universidade de S{\~a}o Paulo, Instituto de F\'{\i}sica, Caixa Postal 66318, S{\~a}o Paulo CEP05315-970, Brazil}
\newcommand{\seoulnat}{System Electronics Laboratory, Seoul National University, Seoul, South Korea}
\newcommand{\stonybrkc}{Chemistry Department, Stony Brook University, Stony Brook, SUNY, NY 11794-3400, U.S.}
\newcommand{\stonycrkp}{Department of Physics and Astronomy, Stony Brook University, SUNY, Stony Brook, NY 11794, U.S.}
\newcommand{\subatech}{SUBATECH (Ecole des Mines de Nantes, CNRS-IN2P3, Universit{\'e} de Nantes) BP 20722 - 44307, Nantes, France}
\newcommand{\tenn}{University of Tennessee, Knoxville, TN 37996, U.S.}
\newcommand{\titech}{Department of Physics, Tokyo Institute of Technology, Oh-okayama, Meguro, Tokyo 152-8551, Japan}
\newcommand{\tsukuba}{Institute of Physics, University of Tsukuba, Tsukuba, Ibaraki 305, Japan}
\newcommand{\vandy}{Vanderbilt University, Nashville, TN 37235, U.S.}
\newcommand{\waseda}{Waseda University, Advanced Research Institute for Science and Engineering, 17 Kikui-cho, Shinjuku-ku, Tokyo 162-0044, Japan}
\newcommand{\weizmann}{Weizmann Institute, Rehovot 76100, Israel}
\newcommand{\yonsei}{Yonsei University, IPAP, Seoul 120-749, Korea}
\affiliation{\abilene}
\affiliation{\banaras}
\affiliation{\bnl}
\affiliation{\caucr}
\affiliation{\charlesczech}
\affiliation{\ciae}
\affiliation{\cns}
\affiliation{\colorado}
\affiliation{\columbia}
\affiliation{\czechtech}
\affiliation{\dapnia}
\affiliation{\debrecen}
\affiliation{\elte}
\affiliation{\fit}
\affiliation{\fsu}
\affiliation{\gsu}
\affiliation{\hiroshima}
\affiliation{\ihepprot}
\affiliation{\illuiuc}
\affiliation{\instpasczech}
\affiliation{\isu}
\affiliation{\jinrdubna}
\affiliation{\kek}
\affiliation{\kfki}
\affiliation{\korea}
\affiliation{\kurchatov}
\affiliation{\kyoto}
\affiliation{\labllr}
\affiliation{\lawllnl}
\affiliation{\losalamos}
\affiliation{\lpc}
\affiliation{\lund}
\affiliation{\muenster}
\affiliation{\myongji}
\affiliation{\nagasaki}
\affiliation{\newmex}
\affiliation{\nmsu}
\affiliation{\ornl}
\affiliation{\orsay}
\affiliation{\peking}
\affiliation{\pnpi}
\affiliation{\riken}
\affiliation{\rikjrbrc}
\affiliation{\rikkyo}
\affiliation{\saispbstu}
\affiliation{\saopaulo}
\affiliation{\seoulnat}
\affiliation{\stonybrkc}
\affiliation{\stonycrkp}
\affiliation{\subatech}
\affiliation{\tenn}
\affiliation{\titech}
\affiliation{\tsukuba}
\affiliation{\vandy}
\affiliation{\waseda}
\affiliation{\weizmann}
\affiliation{\yonsei}
\author{A.~Adare}	\affiliation{\colorado}
\author{S.~Afanasiev}	\affiliation{\jinrdubna}
\author{C.~Aidala}	\affiliation{\columbia}
\author{N.N.~Ajitanand}	\affiliation{\stonybrkc}
\author{Y.~Akiba}	\affiliation{\riken} \affiliation{\rikjrbrc}
\author{H.~Al-Bataineh}	\affiliation{\nmsu}
\author{J.~Alexander}	\affiliation{\stonybrkc}
\author{K.~Aoki}	\affiliation{\kyoto} \affiliation{\riken}
\author{L.~Aphecetche}	\affiliation{\subatech}
\author{R.~Armendariz}	\affiliation{\nmsu}
% ...
\collaboration{PHENIX Collaboration} \noaffiliation

\date{\today}

\begin{abstract}

The momentum distribution of electrons from decays of heavy flavor (charm and beauty) for
midrapidity $|y|<0.35$
in $p + p$ collisions at $\sqrt{s}$~=~200 GeV has been measured by the PHENIX experiment at the Relativistic
Heavy Ion Collider (RHIC) over the transverse momentum range $0.3 < p_{\rm T} < 9$~GeV/$c$. 
Two independent methods have been used to determine the heavy flavor yields,
and the results are in good agreement with each other.
A fixed-order-plus-next-to-leading-log pQCD calculation 
agrees with the data within the theoretical and experimental uncertainties, with the 
data/theory ratio of 
$1.72 \pm~0.02^{\rm stat} \pm~0.19^{\rm sys}$ 
for $0.3 < p_{\rm T} < 9$~GeV/$c$.
The total charm production cross section at this energy has also been deduced to
be $\sigma_{c\bar{c}} = 567 \pm~57^{\rm stat} \pm~224^{\rm sys}~\mu$b.
\end{abstract}


\maketitle

Heavy-flavor (charm and beauty) production serves as a testing
ground of QCD.
Because of the large quark mass,
it is expected that next-to-leading order perturbative QCD
(NLO pQCD) can describe the
production cross section of charm and beauty at high energy,
particularly at high $p_{\rm T}$.
At the Tevatron, beauty production is well described by
NLO pQCD~\cite{Cacciari:2004ur}. Charm production cross sections
at high $p_{\rm T}$ are found to be higher than the theory
by $\approx$ 50 \%, but are compatible 
within the theoretical uncertainties~\cite{Acosta:2003ax}.
Since heavy-flavor production at RHIC energies is dominated by
gluon-gluon fusion, its production in polarized $p+p$
collisions probes the gluon distribution $G(x)$
and the gluon polarization $\Delta G(x)$.
A good understanding of the reaction mechanism for
heavy-flavor production is crucial for reliably extracting
these distributions.
Furthermore in Au+Au collisions at RHIC strong
suppression of single electrons from heavy-flavor decays
has been observed~\cite{Adler:2005xv}.
Measurements of heavy-flavor production
in $p+p$ collisions provide a baseline
for studying hot and dense matter effects in heavy ion reactions.
Earlier measurements at RHIC~\cite{Adler:2005fy,Adams:2004fc} have
a limited $p_{\rm T}$ range with substantial experimental
uncertainties, so an improved measurement is crucial.

We report the production cross section of electrons,
$(e^+ + e^-)/2$, at mid-rapidity in $p+p$ collisions at
$\sqrt{s}$~=~200 GeV for $0.3 < p_{\rm T} < 9$ GeV/$c$
measured by the PHENIX experiment.
Contributions from semi-leptonic decays of heavy-flavor
are determined using two independent methods.
This measurement has over two orders of magnitude larger statistics
with much reduced systematic uncertainties compared to
our previous measurement~\cite{Adler:2005fy}.

The data were collected by the PHENIX detector~\cite{Adcox:2003zm}
during the 2005 RHIC run using the two central arm spectrometers.
Each spectrometer
covers $|\eta| < 0.35$ in pseudo-rapidity and $\Delta \phi = \pi/2$
in azimuth. It includes a drift chamber (DC) and pad chambers (PC1) for
charged particle tracking, a Ring Imaging \v{C}erenkov detector (RICH)
for electron identification, and an electromagnetic calorimeter (EMCal)
for electron identification and trigger. Beam-beam counters
(BBCs), positioned at pseudo-rapidities $3.1< |\eta| < 3.9$, measure
the position of the collision vertex along the beam ($z_{\rm vtx}$) and provide
the interaction trigger.
In this run, helium bags, one for each arm,
were placed in the space between the beam pipe and DC to reduce multiple
scattering and photon conversion.

Two datasets are used for the analysis: (1) the minimum bias (MB) dataset
recorded by the BBC trigger, and (2) a ``photon'' trigger (PH) dataset
triggered at level-1 requiring a minimum energy
deposit of 1.4 GeV in an overlapping tile of $4 \times 4$ EMCal towers
in coincidence with the BBC trigger.
The PH trigger has $\simeq$ 100 \% efficiency for high $p_{\rm T}$ electrons
above 2 GeV/$c$ in the active trigger tiles.
The BBC trigger cross section is
$ 23.0 \pm 2.2$ mb.
Since only $\simeq$ 50 \% of inelastic $p+p$ collisions
satisfy the BBC trigger condition, only a fraction of
the inclusive electron production events is triggered. 
This $p_{\rm T}$ and process independent fraction is
determined to be $\epsilon_{\rm bias} = 0.79 \pm 0.02$
from the yield ratio of high $p_{\rm T}$ $\pi^0$ with and
without the BBC trigger.
After selection of good runs and 
a vertex cut of $|z_{\rm vtx}| < 20$ cm,
an integrated luminosity ($\mathcal{L}$) of
45 nb$^{-1}$ in the MB dataset and 1.57 pb$^{-1}$ 
in the PH dataset are used for the analysis.
%
%

%
Charged particle tracks are reconstructed using
DC and PC1 and confirmed by a hit
in EMCal within 4$\sigma$ in position.
The momentum resolution is
$\sigma_p/p \simeq 0.7 \% \oplus 0.9 \%p$ (GeV/$c$), and
the momentum scale is calibrated within 1\%
using the reconstructed mass of $J/\psi \rightarrow e^+e^-$.

Electron identification (eID) requires at least
two associated hits in RICH, 
a shower shape cut in EMCal, and a cut in the
ratio $E/p$ where $E$ is energy measured in EMCal. 
We require
$0.7 < E/p <1.3$ for $0.8<p_{\rm T}<5$ GeV/$c$. For lower $p_{\rm T}$,
the minimum value of $E/p$ decreases with decreasing
$p_{\rm T}$ to 0.55 at $p_{\rm T}$~=~0.3 GeV/$c$.
The $E/p$ cut removes background electrons from photon conversions
and semi-leptonic decay of kaons ($K\rightarrow e \nu \pi (K_{e3})$)
that occur far from the vertex, and most of the remaining hadron
background.
The hadron contamination after the $E/p$ cut is 3\% at $p_\mathrm{T}=0.3$
GeV/$c$ and less than 1\% for $0.8 < p_{\rm T} < 5$ GeV/$c$
with eID efficiency of approximately 90 \%.

For $p_{\rm T} >$ 5 GeV/$c$, where pions also emit
\v{C}erenkov photons in RICH, tighter electron identification
cuts are applied. We require at least 5 associated hits
in RICH,  a tighter shower shape cut
in EMCal, and $0.8<E/p<1.3$. With these cuts, the electron
measurement is extended to 9 GeV/$c$ in $p_{\rm T}$.
The eID efficiency of the tighter cuts is $p_{\rm T}$ independent, and
is determined to be 57\% of that for $p_{\rm T}<5$ GeV/$c$ by
applying the same tighter cuts for $p_{\rm T}<5$ GeV/$c$.
With the tighter
cuts, hadron contamination is negligible for $p_{\rm T} < 7$ GeV/$c$.
For $7<p_{\rm T}<8$ ($8<p_{\rm T}<9$) GeV/$c$, a 20\% (40\%)
hadron contamination is determined and subtracted
using a Gaussian plus exponential fit to $E/p$ distribution.


The invariant cross section for electron
production is calculated using the following formula,
\begin{equation}
E\frac{d^3\sigma}{dp^3} = \frac{1}{\mathcal{L}}\frac{1}{2\pi p_{\rm T}}\frac{N_e}{\Delta p_{\rm T} \Delta y}\frac{1}{\epsilon_{\rm rec}}\frac{1}{\epsilon_{\rm bias}},
\end{equation}
where $N_e$ is the measured electron yield;
$\epsilon_{\rm rec}$, calculated using a full GEANT~\cite{GEANT} simulation,
includes the geometrical acceptance, track reconstruction and eID efficiency,
and the smearing effect due to finite momentum resolution.
For the PH dataset, $\epsilon_{\rm rec}$ also includes
the PH trigger efficiency. The cross sections from the MB and the PH datasets are
consistent with each other for the overlapped $p_{\rm T}$ region.

The inclusive electron yield consists of three components:
(1) electrons from heavy-flavor decay,
(2) ``photonic'' background electrons from Dalitz decays
of light mesons and photon conversions primarily in the
beam pipe, and
(3)``non-photonic'' background electrons
from the remaining $K_{e3}$ decays and dielectron decays of vector mesons.
The photonic background is much larger than the non-photonic background.
We determined the spectrum of electrons from heavy-flavor
decay by subtracting the background components from the
inclusive spectrum using the following two independent
methods.

In the ``cocktail subtraction'' method~\cite{Adler:2005fy,Adler:2005xv,Adcox:2002cg}
a cocktail of electron spectra from various background sources is
calculated using a Monte Carlo event generator of hadron decays. 
The most important background is the $\pi^0$ Dalitz decay, so we use
our measured $\pi^0$ and $\pi^{\pm}$ spectra as input to the generator.
The spectral shapes of other
light hadrons $h$ are obtained from the pion spectra by
$m_T$ scaling ($p_{\rm T} \rightarrow \sqrt{p_{\rm T}^2+M_h^2-M_\pi^2})$.
Within this approach the ratios $h/\pi^0$ are constant
at high $p_{\rm T}$. For the relative normalization, we use the
following ratios:
$\eta/\pi^0=0.48 \pm 0.03$~\cite{Adler:2006hu},
$\rho^0/\pi^0=1.0 \pm 0.3$,
$\omega/\pi^0=0.90 \pm 0.06$~\cite{Ryabov:2005xv},
$\eta'/\pi^0=0.40 \pm 0.12$,
$\phi/\pi^0=0.25 \pm 0.08$.
For $p_{\rm T} >2$ GeV/$c$, contributions from $\eta$ and
all other hadrons combined are approximately
20\% and 10\% of $\pi^0$, respectively.
Another major background electron source is
conversions of photons in the beam pipe
(0.29\% of a radiation length($X_0$)) as well as
in the air and the helium bags (0.1\% $X_0$).
The conversion electron spectrum is very similar to that of
Dalitz decays. Using a detailed GEANT
simulation of the PHENIX detector, the ratio of
electrons from conversions to Dalitz decays,
$R_\mathrm{CD}$, is determined to be 
0.40 $\pm$ 0.04 for
$\pi^0$, essentially $p_{\rm T}$ independent. 
$R_\mathrm{CD}$ is approximately half of that in~\cite{Adler:2005fy}
since the helium bags eliminated most of the conversions
outside of the beam pipe.
The conversion spectra are calculated by
scaling the Dalitz decay spectra by $R_\mathrm{CD}$,
with small corrections to account for the species dependence of the
relative branching ratio of Dalitz decay to photon decay
($(h\rightarrow ee\gamma)/(h\rightarrow \gamma\gamma)$).
The internal and external conversions of direct photons are also included in the
cocktail, using our measured direct photon spectrum~\cite{PPG060}
as input.
The direct photon contribution is comparable to or greater than
that from the $\eta$ for $p_{\rm T}>5$ GeV/$c$.
Non-photonic backgrounds are also included in the cocktail.
Since the $K_{e3}$ background depends on the analysis cuts, it
is evaluated by a full GEANT simulation.

In the ``converter subtraction'' method~\cite{Adler:2004ta},
we introduce an additional photon converter (a thin brass sheet
of 1.67\% $X_0$) around the beam pipe for part of the run.
The converter multiplies the photonic electron background by a fixed factor,
$R_{\gamma} \simeq 2.3$, which
is determined precisely via GEANT simulation.
$R_\gamma$ is larger than in~\cite{Adler:2004ta}
since we have less conversion material in the 2005 run.
The photonic background $N_{e}^{\gamma}$ is determined
as $N_{e}^{\gamma} = (N_e^{\rm C}-(1-\epsilon)N_e^{\rm NC})/(R_\gamma-1+\epsilon)$,
where $N_{e}^{\rm C}$ and $N_{e}^{\rm NC}$ are electron yield with and without
the converter, respectively; and
$\epsilon$ (2.1\%) represents a small loss of electrons
due to the converter. The non-photonic component is then
determined as $N_e^{{\rm non}-\gamma}=N_e^{\rm NC}-N_e^{\gamma}$.
Small remaining non-photonic background, such as
$K_{e3}$ and hadron contamination, are subtracted.

These two methods are complementary to each other.
The converter method is more accurate, and it allows us to
extract a heavy-flavor signal down to
$p_{\rm T} = 0.3$ GeV/$c$ where the signal is only
$\approx$ 10\% of inclusive electrons.
In addition, the measured photonic background $N_e^{\gamma}$
is used to confirm and to calibrate the normalization of
the calculated cocktail yields.
A drawback of the method is statistical precision: the converter run contains
only a small fraction ($\simeq 7\%$ in the 2005 run) of the data.
The cocktail method can use the full statistics at high $p_{\rm T}$, where
the photonic background becomes a small fraction of inclusive electrons.

Systematic uncertainties are
categorized into (a) inclusive electron spectra, (b)
cocktail subtraction, and (c) converter
subtraction. Category (a) is common to both analyses,
and includes the uncertainties in luminosity (9.6\%),
geometrical acceptance (4\%), eID efficiency (3\%),
and the PH trigger efficiency (3\% at the plateau).
Uncertainties in cocktail subtraction (category (b))
include the normalization (8\%)
and $p_{\rm T}$ dependent shape uncertainty
(2\% at $p_{\rm T} \simeq$ 2 GeV/$c$, increasing to
6\% at 9 GeV/$c$).
In the converter analysis (category(c)) the dominant
uncertainties are in $R_\gamma$ (2.7\%) and in
the relative acceptance in the converter and the normal
runs (1.0\%).
These uncertainties are propagated into the
uncertainties in the heavy-flavor electron yields and added in
quadrature.

\begin{figure}[tb]
\includegraphics[width=1.0\linewidth]{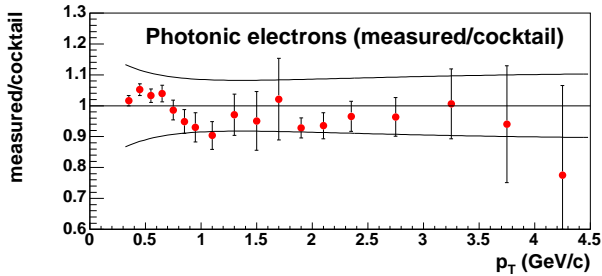}
\caption{\label{fig:1}
Ratio of photonic electrons measured by the
converter method to the cocktail calculation.
Data from the MB (PH) dataset are shown below
(above) 1.8 GeV/$c$. The upper and lower curves show the
systematic error of the cocktail. Error bars are statistical only.
}
\end{figure}

Figure~\ref{fig:1} shows the ratio of the measured
$N_{e}^{\gamma}$ to the cocktail calculation as a function
of $p_{\rm T}$. The ratio is consistent with unity within the
uncertainties of the cocktail. At high $p_{\rm T}$ ($> 1.8$ GeV/$c$),
the ratio is $0.94 \pm 0.02^{\rm stat}$ on average.
Since this is within the uncertainty of the cocktail normalization,
we rescale the cocktail yields by
this factor. This removes the 8\% normalization uncertainty in the
cocktail.

\begin{figure}[tb]
\includegraphics[width=1.0\linewidth]{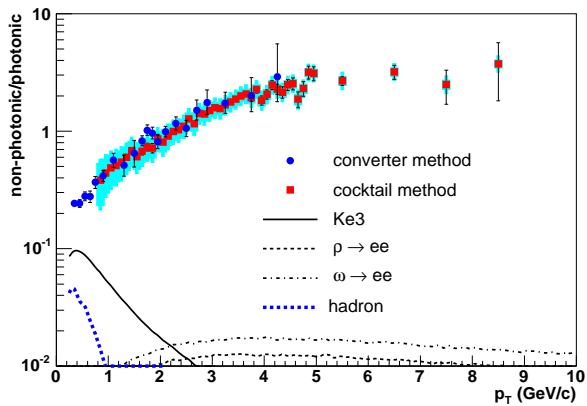}
\caption{\label{fig:2} 
Ratio of non-photonic electrons to photonic background.
Error bars are statistical errors and the error bands
show the cocktail systematic errors.
The solid, dashed, dot-dashed, and dot curve is
the remaining non-photonic background from $K_{e3}$,
$\rho \rightarrow ee$,
$\omega \rightarrow ee$, and hadron contamination, respectively.
}
\end{figure}

In Fig.~\ref{fig:2}, filled circles (squares) show the
ratio of non-photonic electrons relative to
photonic background determined by the converter
(cocktail) method.
The non-photonic electrons
are dominantly heavy-flavor decay signals.
The remaining non-photonic background contributions
have been calculated and are shown in Fig.~\ref{fig:2}.
The two methods are consistent with each other.
The ratio monotonically increases with increasing $p_{\rm T}$,
becoming greater than unity for $p_{\rm T} >$ 2.4 GeV/$c$, 
and saturates at $\simeq 3$ for $p_{\rm T} > 5$ GeV/$c$.
The large signal-to-background ratio is due to the small amount
of conversion material in the spectrometer acceptance.

\begin{figure}[tb]
\includegraphics[width=1.0\linewidth]{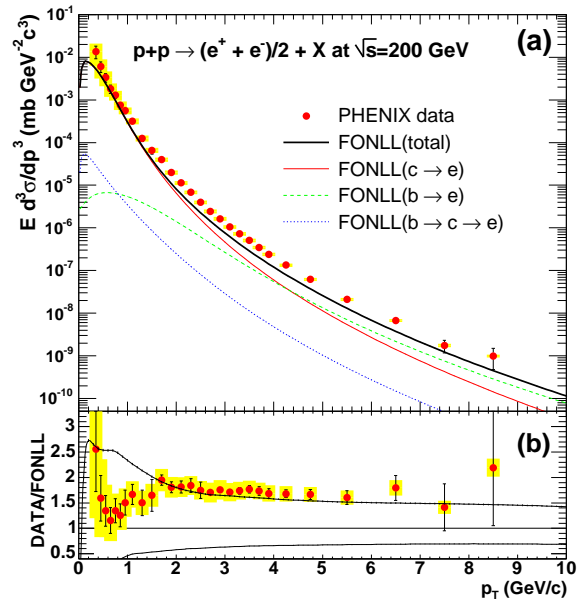}
\caption{\label{fig:3}
(a) Invariant differential cross sections of electrons from
heavy-flavor decays. The error bars (bands) represent the
statistical (systematic) errors.
The curves are the FONLL calculations (see text).
(b) Ratio of the data and the FONLL calculation. The upper
(lower) curve shows the theoretical upper (lower) limit
of the FONLL calculation. In both panels a 10\% normalization
uncertainty is not shown.
}
\end{figure}

Figure~\ref{fig:3}~(a) shows the invariant differential cross section
of electrons from heavy-flavor decays.
All background has been subtracted, including 
the non-photonic components shown in Fig.~\ref{fig:2}.
The data from the two analysis 
methods are combined:
at low $p_{\rm T}$ ($p_{\rm T}$ $< 1.6$ GeV/$c$) the converter subtraction method on
the MB data set is used; at intermediate
$p_{\rm T}$ ($1.6< p_{\rm T} < 2.6$~GeV/$c$) the converter method on the PH dataset
is used; and at high $p_{\rm T}$ ($p_{\rm T}>2.6$~GeV/$c$) the cocktail method on the PH dataset
is used.

The data are compared with a fixed-order-plus-next-to-leading-log
(FONLL) pQCD calculation~\cite{Cacciari:2005rk,Cacciari:pc}.
The top curve in Fig.~\ref{fig:3} shows the central values of
the FONLL calculation. The contributions of charm and beauty
are also shown. For $p_{\rm T} > 4$ GeV/$c$, the beauty contribution becomes
dominant. In Fig.~\ref{fig:3}~(b), the ratio of the data
to the FONLL calculation
is shown. The ratio is nearly $p_{\rm T}$ independent over the entire
$p_{\rm T}$ range. Fitting to a constant for $0.3< p_{\rm T} <9.0$ GeV/$c$ yields a ratio of
$1.72 \pm~0.02^{\rm stat} \pm~0.19^{\rm sys}$ 
Similar ratios are observed in charm production
at high $p_{\rm T}$ at the Tevatron~\cite{Acosta:2003ax}.
The upper limit of the FONLL calculation
is compatible with the data.
Recently STAR reported~\cite{Abelev:2006db} that non-photonic electron
production in $p+p$ at $\sqrt{s}=200$ GeV
is 5.5 times larger than predicted by the same FONLL calculation.
We do not observe such a large discrepancy.
We note that the photonic electron background in our spectrometer
is approximately 1/10 that of STAR due to the small amount of conversion material
in the PHENIX acceptance.

%
%
%
The total charm cross section is derived by
integrating the heavy-flavor electron cross section for $p_{\rm T}>$ 0.4 GeV/$c$:
$d\sigma_e(p_{\rm T}>0.4)/dy = 5.95 \pm~0.59 \pm~2.0~\mu$b.
The systematic error is obtained by integrating the upper and lower
systematic error limits of the differential cross sections, since
the systematic errors are essentially
coherent.
The cross section is then
extrapolated to $p_{\rm T}$ = 0 using the spectrum
shape predicted by FONLL:
$d\sigma_e(p_{\rm T}>0)/dy=10.9 \pm 1.1 \pm 3.8~\mu$b.
We have assigned 10\% to the systematic
uncertainty of the extrapolation, and have subtracted contribution
from beauty and beauty cascade decays ($0.1 \mu$b).
We determine the charm production cross section,
$d\sigma_{c\bar{c}}/dy$ = $123 \pm 12 \pm 45~\mu$b, by
using a $c \rightarrow e$ total branching ratio of $9.5 \pm 1.0$\%,
calculated using the following charmed hadron ratios:
$D^+/D^0 = 0.45 \pm 0.1, D_s/D^0 = 0.25 \pm 0.1$, and
$\Lambda_c/D^0 = 0.1 \pm 0.05$. The rapidity distribution of electrons is
broader than that of $D$ mesons due to decay kinematics.
A correction to this effect (7\%) has been applied.
Using the rapidity distribution from HVQMNR~\cite{Mangano:1992kq} with
CTEQ5M~\cite{Lai:1999wy} PDF, the total
charm cross section is determined to 
be $\sigma_{c\bar{c}} = 567 \pm~57^{\rm stat} \pm~224^{\rm sys}~\mu$b.
We have assigned 15\% systematic error to the extrapolation.
This result is compatible with
our previous measurement~\cite{Adler:2005fy} ($920 \pm 150 \pm 540~\mu$b)
and the value derived from Au+Au
collisions~\cite{Adler:2004ta} ($622 \pm 57 \pm 160~\mu$b per $NN$
collisions).
The FONLL cross section
($256^{+400}_{-146}~\mu$b) is compatible
with the data within its
uncertainty. STAR has reported a somewhat larger value
in $d$+Au~\cite{Adams:2004fc} ($1.3~\pm~0.2~\pm~0.4$~mb per $NN$ collisions).
Although the data extend to high $p_{\rm T}$  where the beauty
contribution is expected to be dominant,
the present analysis does not separate charm and beauty
contributions. The beauty cross section predicted by
FONLL is $1.87^{+0.99}_{-0.67}~\mu$b, and the upper FONLL
curve is consistent with the data.

In conclusion, we have measured single electrons from
heavy-flavor decays in $p+p$ collisions at
$\sqrt{s}$~=~200 GeV.
The new data reported here provide a crucial
baseline for the study of heavy quark
production in hot and dense matter created in
Au+Au collisions.
The agreement between the data and the FONLL pQCD calculation
within the theoretical and the experimental uncertainties
suggests that a reliable extraction of gluon polarization
from heavy-flavor production in 
polarized $p+p$ collisions is attainable.

We thank the staff of the Collider-Accelerator and Physics Departments at BNL
for their vital contributions.
We acknowledge support from the Department of Energy and NSF (U.S.A.),
MEXT and JSPS (Japan), CNPq and FAPESP (Brazil), NSFC (China),
MSMT (Czech Republic), IN2P3/CNRS, and CEA (France),
BMBF, DAAD, and AvH (Germany), OTKA (Hungary), DAE (India),
ISF (Israel), KRF and KOSEF (Korea), MES, RAS, and FAAE (Russia),
VR and KAW (Sweden), U.S. CRDF for the FSU, US-Hungarian
NSF-OTKA-MTA, and US-Israel BSF.

\def\IJMPA{{Int. J. Mod. Phys.}~{\bf A}}
\def\JPG{{J. Phys}~{\bf G}}
\def\NCA{Nuovo Cimento}
\def\NIM{Nucl. Instrum. Methods}
\def\NIMA{{Nucl. Instrum. Methods}~{\bf A}}
\def\NPA{{Nucl. Phys.}~{\bf A}}
\def\NPB{{Nucl. Phys.}~{\bf B}}
\def\PLB{Phys. Lett. B}
\def\PLC{Phys. Repts.\ }
\def\PRL{Phys. Rev. Lett.\ }
\def\PRD{Phys. Rev. D}
\def\PRC{Phys. Rev. C}
\def\ZPC{{Z. Phys.}~{\bf C}}
\def\etal{{\it et al.}}


\begin{thebibliography}{17}
\expandafter\ifx\csname natexlab\endcsname\relax\def\natexlab#1{#1}\fi
\expandafter\ifx\csname bibnamefont\endcsname\relax
  \def\bibnamefont#1{#1}\fi
\expandafter\ifx\csname bibfnamefont\endcsname\relax
  \def\bibfnamefont#1{#1}\fi
\expandafter\ifx\csname citenamefont\endcsname\relax
  \def\citenamefont#1{#1}\fi
\expandafter\ifx\csname url\endcsname\relax
  \def\url#1{\texttt{#1}}\fi
\expandafter\ifx\csname urlprefix\endcsname\relax\def\urlprefix{URL }\fi
\providecommand{\bibinfo}[2]{#2}
\providecommand{\eprint}[2][]{\url{#2}}

\bibitem[{\citenamefont{Cacciari}(2004)}]{Cacciari:2004ur}
\bibinfo{author}{\bibfnamefont{M.}~\bibnamefont{Cacciari}}
  (\bibinfo{year}{2004}), \eprint{hep-ph/0407187}.

\bibitem[{\citenamefont{Acosta et~al.}(2003)}]{Acosta:2003ax}
\bibinfo{author}{\bibfnamefont{D.}~\bibnamefont{Acosta}} \bibnamefont{et~al.},
  \bibinfo{journal}{Phys. Rev. Lett.} \textbf{\bibinfo{volume}{91}},
  \bibinfo{pages}{241804} (\bibinfo{year}{2003}).

\bibitem[{\citenamefont{Adler et~al.}(2006{\natexlab{a}})}]{Adler:2005xv}
\bibinfo{author}{\bibfnamefont{S.~S.} \bibnamefont{Adler}}
  \bibnamefont{et~al.}, \bibinfo{journal}{Phys. Rev. Lett.}
  \textbf{\bibinfo{volume}{96}}, \bibinfo{pages}{032301}
  (\bibinfo{year}{2006}{\natexlab{a}}).

\bibitem[{\citenamefont{Adler et~al.}(2006{\natexlab{b}})}]{Adler:2005fy}
\bibinfo{author}{\bibfnamefont{S.~S.} \bibnamefont{Adler}}
  \bibnamefont{et~al.}, \bibinfo{journal}{Phys. Rev. Lett.}
  \textbf{\bibinfo{volume}{96}}, \bibinfo{pages}{032001}
  (\bibinfo{year}{2006}{\natexlab{b}}).

\bibitem[{\citenamefont{Adams et~al.}(2005)}]{Adams:2004fc}
\bibinfo{author}{\bibfnamefont{J.}~\bibnamefont{Adams}} \bibnamefont{et~al.},
  \bibinfo{journal}{Phys. Rev. Lett.} \textbf{\bibinfo{volume}{94}},
  \bibinfo{pages}{062301} (\bibinfo{year}{2005}).

\bibitem[{\citenamefont{Adcox et~al.}(2003)}]{Adcox:2003zm}
\bibinfo{author}{\bibfnamefont{K.}~\bibnamefont{Adcox}} \bibnamefont{et~al.},
  \bibinfo{journal}{Nucl. Instrum. Meth.} \textbf{\bibinfo{volume}{A499}},
  \bibinfo{pages}{469} (\bibinfo{year}{2003}).

\bibitem[{GEA()}]{GEANT}
\bibinfo{note}{{CERN Program Library}}.

\bibitem[{\citenamefont{Adcox et~al.}(2002)}]{Adcox:2002cg}
\bibinfo{author}{\bibfnamefont{K.}~\bibnamefont{Adcox}} \bibnamefont{et~al.},
  \bibinfo{journal}{Phys. Rev. Lett.} \textbf{\bibinfo{volume}{88}},
  \bibinfo{pages}{192303} (\bibinfo{year}{2002}).

\bibitem[{\citenamefont{Adler et~al.}(2006{\natexlab{c}})}]{Adler:2006hu}
\bibinfo{author}{\bibfnamefont{S.~S.} \bibnamefont{Adler}}
  \bibnamefont{et~al.}, \bibinfo{journal}{Phys. Rev. Lett.}
  \textbf{\bibinfo{volume}{96}}, \bibinfo{pages}{202301}
  (\bibinfo{year}{2006}{\natexlab{c}}).

\bibitem[{\citenamefont{Ryabov et~al.}(2005)}]{Ryabov:2005xv}
\bibinfo{author}{\bibfnamefont{V.}~\bibnamefont{Ryabov}} \bibnamefont{et~al.}
  (\bibinfo{year}{2005}), \eprint{hep-ex/0510017}.

\bibitem[{\citenamefont{Adler et~al.}()}]{PPG060}
\bibinfo{author}{\bibfnamefont{S.~S.} \bibnamefont{Adler}}
  \bibnamefont{et~al.}, \bibinfo{note}{to be published}.

\bibitem[{\citenamefont{Adler et~al.}(2005)}]{Adler:2004ta}
\bibinfo{author}{\bibfnamefont{S.~S.} \bibnamefont{Adler}}
  \bibnamefont{et~al.}, \bibinfo{journal}{Phys. Rev. Lett.}
  \textbf{\bibinfo{volume}{94}}, \bibinfo{pages}{082301}
  (\bibinfo{year}{2005}).

\bibitem[{\citenamefont{Cacciari et~al.}(2005)}]{Cacciari:2005rk}
\bibinfo{author}{\bibfnamefont{M.}~\bibnamefont{Cacciari}}
  \bibnamefont{et~al.}, \bibinfo{journal}{Phys. Rev. Lett.}
  \textbf{\bibinfo{volume}{95}}, \bibinfo{pages}{122001}
  (\bibinfo{year}{2005}).

\bibitem[{\citenamefont{Cacciari}()}]{Cacciari:pc}
\bibinfo{author}{\bibfnamefont{M.}~\bibnamefont{Cacciari}},
  \bibinfo{note}{private communication}.

\bibitem[{\citenamefont{Abelev}(2006)}]{Abelev:2006db}
\bibinfo{author}{\bibfnamefont{B.~I.} \bibnamefont{Abelev}}
  (\bibinfo{year}{2006}), \eprint{nucl-ex/0607012}.

\bibitem[{\citenamefont{Mangano et~al.}(1993)\citenamefont{Mangano, Nason, and
  Ridolfi}}]{Mangano:1992kq}
\bibinfo{author}{\bibfnamefont{M.~L.} \bibnamefont{Mangano}},
  \bibinfo{author}{\bibfnamefont{P.}~\bibnamefont{Nason}}, \bibnamefont{and}
  \bibinfo{author}{\bibfnamefont{G.}~\bibnamefont{Ridolfi}},
  \bibinfo{journal}{Nucl. Phys.} \textbf{\bibinfo{volume}{B405}},
  \bibinfo{pages}{507} (\bibinfo{year}{1993}).

\bibitem[{\citenamefont{Lai et~al.}(2000)}]{Lai:1999wy}
\bibinfo{author}{\bibfnamefont{H.~L.} \bibnamefont{Lai}} \bibnamefont{et~al.},
  \bibinfo{journal}{Eur. Phys. J.} \textbf{\bibinfo{volume}{C12}},
  \bibinfo{pages}{375} (\bibinfo{year}{2000}).

\end{thebibliography}

\end{document}